\RequirePackage{fix-cm}
\documentclass[twocolumn]{svjour3}  % twocolumn

\usepackage{dblfloatfix}
\usepackage{subfig}
\usepackage{booktabs}
\smartqed  % flush right qed marks, e.g. at end of proof
\usepackage{graphicx}
\usepackage[numbers]{natbib}

\journalname{Universal Access and Information Society}
\begin{document}

% BibTeX users please use one of
%\bibliographystyle{spbasic}      % basic style, author-year citations
%\bibliographystyle{spmpsci}      % mathematics and physical sciences
%\bibliographystyle{spphys}       % APS-like style for physics
%\bibliography{}   % name your BibTeX data base
%\bibliography{sample-base.bib}

\title{Supporting a Crowd-powered Accessible Online Art Gallery for People with Visual Impairments: A Feasibility Study%\thanks{Grants or other notes
%about the article that should go on the front page should be
%placed here. General acknowledgments should be placed at the end of the article.}
}
%\subtitle{Do you have a subtitle?\\ If so, write it here}

%\titlerunning{Short form of title}        % if too long for running head

%\author{Anonymous Submission}
\author{{Nahyun Kwon}$^{1}$ \and
        {Yunjung Lee}$^{1}$ \and
        {Uran Oh}$^{1}$%etc.
}

%\authorrunning{Short form of author list} % if too long for running head

\institute{Uran Oh, the corresponding author \at
              %first address \\
              %Tel.: +123-45-678910\\
              %Fax: +123-45-678910\\
              \email{uran.oh@ewha.ac.kr}           %  \\
              %\emph{Present address:} of F. Author  %  if needed
           \and
           $^{1}$ Department of Computer Science and Engineering, Ewha Womans University, Republic of Korea
}

\date{Received: date / Accepted: date}
% The correct dates will be entered by the editor

\titlerunning{Accessible Online Art Gallery}
\maketitle

\begin{abstract}
While people with visual impairments are interested in artwork as much as their sighted peers, their experience is limited to few selective artworks that are exhibited at certain museums. 
To enable people with visual impairments to access and appreciate as many artworks as possible at ease, we propose an online art gallery that allows users to explore different parts of a painting displayed on their touchscreen-based devices while listening to corresponding verbal descriptions of the touched part on the screen. To investigate the scalability of our approach, we first explored if anonymous crowd who may not have expertise in art are capable of providing visual descriptions of artwork as a preliminary study. Then we conducted a user study with 9 participants with visual impairments to explore the potential of our system for independent artwork appreciation by assessing if and how well the system supports 4 steps of Feldman Model of Criticism. The findings suggest that visual descriptions of artworks produced by an anonymous crowd are sufficient for people with visual impairments to interpret and appreciate paintings with their own judgments which is different from existing approaches that focused on delivering descriptions and opinions written by art experts. Based on the lessons learned from the study, we plan to collect visual descriptions of a greater number of artwork and distribute our online art gallery publicly to make more paintings accessible for people with visual impairments. 

%\keywords{First keyword \and Second keyword \and More}

% \PACS{PACS code1 \and PACS code2 \and more}
% \subclass{MSC code1 \and MSC code2 \and more}
\end{abstract}

\section{Introduction}

People with visual impairments (PVI) are interested in enjoying artworks as much as sighted people \cite{handa2010investigation, hayhoe2013expanding}. However, they often face accessibility issues when appreciating artworks. For example, most of accessible artworks require them to physically visit certain museums which provides accessible art \cite{asakawa2018present}. 
%because they are inevitably dependent to the help of sighted people. 
Although a number of museums have been providing audio descriptions of exhibited artworks to visitors, they rarely considered the fundamental requirements of PVI such as a detailed visual information of the artworks \cite{rector2017eyes}. 

To better support PVI to appreciate artwork, a number of studies have been conducted \cite{cantoni2018art,iranzo2019exploring, rector2017eyes, OutLoud}. For instance, Eyes-Free Art \cite{rector2017eyes} was proposed to provide PVI with on-site assistance at a museum with various types of audio feedback depending on the distance between a user and the painting that is in front of the user. Iranzo Bartolome \textit{et al.} \cite{iranzo2019exploring} also investigated 3D-printed multi-modal paintings, including both audio and tactile feedback to enable PVI to explore and listen to paintings by touch. While promising, experiences with these existing approaches are limited to a very few selective custom-made artwork replicas that can only be accessed at certain physical exhibition sites. To provide easy access to relatively a greater number of artwork as possible for PVI, another prior study by Kwon and Oh \cite{10.1145/3308561.3354620} has proposed a web application for touchscreen devices that allows PVI to explore a selected painting by dragging their finger to different objects of the painting displayed on a screen while listening to verbal descriptions of the touched locations. %as explored in prior studies \cite{seeingAI, iranzo2019exploring, morris2018rich, 10.1145/3308561.3354620}. 
Although their ultimate goal was to enable PVIs to explore a variety of paintings with object-level descriptions without having to visit certain locations for exploring accessible replicas of the limited number of artwork, collecting the required information for individual paintings is not scalable with a small number of art experts. In addition, it is unclear whether the existing approaches of conveying encyclopedic and/or visual knowledge of artwork can lead PVI to genuine appreciation of artwork. 

Thus, this work aims to improve the accessibility of a greater number of artwork for people with visual impairments by enabling them to access artworks online with their touchscreen devices. Our key research questions (RQ's) include:
%\begin{quote}
\begin{itemize}
    %\item \textit{RQ1. How can we collect visual descriptions for artwork as many as possible?}
    \item \textit{RQ1. How can we increase access to a greater number of various artwork for people with visual impairments?}
    \item \textit{RQ2. To what extent, can touchscreen-based artwork exploration support independent artwork appreciation process for people with visual impairments?}
\end{itemize}
%\end{quote}

To address the first question, we focused on collecting object-level artwork descriptions from people who do not necessarily have art-related expertise. Inspired by WikiArt\footnote{https://www.wikiart.org} which is a visual art encyclopedia following the Wikipedia model\footnote{https://www.wikipedia.org} powered by collective intelligence \cite{lichtenstein2009wikipedia},%where anyone can contribute to refine information of artwork such as translating texts into other languages and reporting low quality images. 
%as it is not scalable to collect the required information by a couple of experts in art. 
we asked anonymous crowd, Amazon Mechanical Turk\footnote{https://www.mturk.com/} (MTurk) workers, to perform annotation tasks of 8 different paintings and analyzed the results to assess the feasibility of this approach. 
%whether the descriptions collected from people who do not necessarily be experts in art is as useful and informative compared to those collected by experts.
%The findings suggests the feasibility of involving crowd for generating visual descriptions for supporting various artwork as many as possible 
%scalable얘기. 
%As a result, our goal is to design a scalable system that can improve painting accessibility for PVI which enables them to explore various paintings by touch on a personal touchscreen device. 
%To design a system that enables PVI to appreciate a variety of artwork, 
%We evaluated our application's ability based on \textit{Feldman Model of Criticism}, which is one of the ways to help PVI to understand and to improve their abilities in appreciating artwork by considering artwork' cultural and social values. 
%[여기에 다양한 작품을 지원하기위해서 크라우드소싱을 사용했고, 위키피디아의 집단 지성 아이디어를 착안해서 누구나 수정가능하다라는 내용 넣기 ]
 %   \setlength{\itemindent}{-0em}
%\textit{RQ1.} How feasible is it to provide information of artwork in detail via data annotated by crowds? 
%In what order, and which level of details about objects within an artwork should be provided to PVI?
%
As for answering the second question, we designed and implemented a prototype based on Shneiderman’s Visual Information-Seeking Mantra \cite{shneiderman1996eyes} (\textit{i.e.}, \textit{overview first}, \textit{zoom \& filter}, \textit{details on demand}). We then conducted a user study with 9 PVI where participants were asked to explore artworks using our prototypes and answer a set of the representative questions that are asked for each of the four steps in \textit{Feldman Model of Criticism}~\cite{PracticalArtCriticism} (\textit{i.e.}, \textit{description}, \textit{analysis}, \textit{interpretation}, \textit{judgment}), which is widely used for practicing artwork appreciation~\cite{clements1979inductive, feldman1987varieties, perkinsArts}. 

%We then designed and implemented a prototype system based on Shneiderman’s Visual Information-Seeking Mantra \cite{shneiderman1996eyes} and conducted a user study with 9 PVI where they were asked to explore two paintings using our system; one with descriptions provided by experts and another one with descriptions generated by mTurk workers. 

The findings of two studies suggest that an anonymous crowd who may or may not have expert knowledge in art is capable of producing visual descriptions of paintings and that our system enables PVI to appreciate and interpret paintings independently as opposed to passively listening to other's descriptions and opinions without having their own judgment and appreciation. 
%to understand needs and preferences of PVI and to design interactions %of an educational tool for appreciating artwork, 

The key contributions of this paper are as follows: (1) a confirmation of the feasibility of collecting visual descriptions of artworks from an anonymous crowd, (2) a proposal and implementation of a system that enables people with visual impairments to appreciate various artwork independently on their personal touchscreen-based devices, (3) the qualitative assessments of our system in terms of Feldman Model of Criticism and (4) design implications for supporting an accessible online art gallery for people with visual impairments.  

\section{Related Work}
Our study builds upon prior studies in the areas of image accessibility for PVI focusing on artworks. 
%Understanding and appreciating works of art is possible using a universal touchscreen and trying to lower barriers.

\subsection{Making Images Accessible on Web}
Various studies have worked on improving the accessibility of visual information of images on the web for PVI by providing alternative text (also called alt text) \cite{10.1145/3308561.3354629, seeingAI, stangl2018browsewithme, winters2019strategies, zhong2018identification}.  
%It is Screen readers cannot render an image as audio unless an alternative text (also called alt text) is specified for that image. However, the automation of alternative text is not desirable. It is because there are several types of images that are difficult to automatically assess the importance. 
For instance, Zhong \textit{et al.} applied the Human-in-the-Loop approach to identify images on a web page that are worth generating alt texts unlike decorative images \cite{zhong2018identification}. Winters \textit{et al.} \cite{winters2019strategies} also proposed an auditory display for social media platform. It provides non-speech auditory icons describing the genders and expressions of any faces in a photo while playing nonvocal background music to convey the overall mood to people who use screen readers. Twitter A11y \cite{10.1145/3308561.3354629} suggested an end-to-end system that creates and retrieves the alt text for Twitter images using a browser extension and crowdsourcing. 
BrowseWithMe \cite{stangl2018browsewithme}, which is a screen reader system specific for shopping websites for clothes, allows PVI to query for information of a product such as price and material using voice input based on the content extracted from a source code of a specific shopping website using its natural language processing module and uses image processing and computer vision techniques to automatically generate a description of the entire outfit given a product image. 
Similarly, Morris \textit{et al.} \cite{morris2018rich} proposed a mobile interface that provides screen reader users rich information of visual contents prepared using real-time crowdsourcing and friend-sourcing; users can listen to not only the alt text of an entire photo but also can drag their fingers over different areas of a photo to listen to information specific to the region that is being touch and ask questions using voice input.  
EyeDescribe \cite{Reinholt:2019:ECE:3343055.3359722} suggested an image labeling system that combines an eye gaze data and a spoken description of the gazed object. It enables the user to explore a spatially annotated image using a touchscreen based application.
%it can provide a new experience of visual content for screen reader users. 
%six different properties of this taxonomy: progressive exploration, multimedia, spatial, categories, question \& answer, and hyperlink~\cite{morris2018rich}. B
While these systems relied on automatic approaches to generate visual attributes of various types of images that PVI may encounter while performing web-based tasks (\textit{i.e.}, shopping for clothes and browsing photos posted on social media), automating the same process for artwork is challenging compared to photos of actual products, people or scenes since the visual appearance of the same figure can be different across paintings due to various artists' expression techniques or materials used for drawing. On the other hand, there is a limited number of masterpieces that we can focus on generating alt texts whereas the number of images on the web increases rapidly. For this reason, we focused on collecting visual descriptions of artwork from an an anonymous crowd and explore the feasibility of this approach as the descriptions can be less objective although we expect that the accuracy of automatic segmentation of objects in paintings to improve in the near feature. Goncu and Marriott \cite{goncu2015creating} already tested creating accessible images by the general public using GraAuthor which is a web-based tool that enables sighted people to create accessible graphics for PVI.  
Moreover, we also applied the approach of playing location-specific descriptions depending on the touched location on touchscreen-based devices used in a prior work \cite{morris2018rich} to our proposed system so that PVI can understand the components of paintings as well as locations of specific objects in paintings. 

%Our study builds on prior studies in the areas of image accessibility for visually impaired people focusing on art paintings. 
\subsection{Making Artwork Accessible for PVI}
 There are studies proposing various solutions for helping PVI to understand and appreciate artwork. Several museums, for instance, offer audio guides that are accessible to anyone including people who are blind or have visual impairments that play verbal descriptions of exhibited artwork \cite{Metropolitan, MOMA, SAMA, OutLoud}. Similarly, Asakawa \textit{et al.}~\cite{guerreiro2019independent} also proposed a mobile application that plays audio content of artwork if located closely in front of a user while helping PVI to navigate a museum with its indoor navigation capability. Eyes-Free Art \cite{rector2017eyes} is another example that allows PVI to access different types of audio feedback based on the distance to a specific 2D painting including a sound effect of an object (\textit{e.g.}, a breeze flowing through the grass for a landscape). 
 However, these require PVI to visit certain exhibition sites in person which might not be ideal considering the survey results conducted by Asakawa \cite{guerreiro2019independent} that almost three quarters of 19 PVI had never been to a museum independently. 
 Making Sense of Art \cite{Holloway:2019:MSA:3290605.3300250} also suggested a new model consists of 3 different ways that enable an independent artwork appreciation of PVI. 
 Moreover, tactile interfaces were proposed to improve image accessibility \cite{gyoshev2018exploiting,cantoni2018art,iranzo2019exploring, cavazos2018interactive}. Cantoni \textit{et al.}, for instance, proposed a model that performs salient segmentation of 2D art masterpieces to extract the contours of each object segment for 3D printing \cite{cantoni2018art}. A multi-modal 2.5D tactile representations of artwork with a voice command capability was also proposed \cite{iranzo2019exploring}; a user can ask for specific information about a 3D printed 2.5D painting such as the title or the name of the artist. While most approaches of representing artwork for PVI focused on 2D paintings, \textit{Touching Masterpieces} \cite{TouchingMP} created 3D models of famous sculptures which can be downloaded and experienced in a virtual environment with a pair of haptic gloves so that PVI can understand the shape and the size of the sculptures with vibration feedback from the gloves. %Regardless of feedback channels, PVI have to be at a museum. Moreover, often not all artwork are made accessible even if they are on-site. More artwork can be reached by more number of PVI via 3D printers, tactile displays or virtual reality devices. However, limitations still exist in accessing these special devices. Our study used mobile devices with touch screens (\textit{e.g.}, iPads), so that users can access and manipulate the device more conveniently. 
 Rodrigues \textit{et al.} \cite{rodrigues2018image} also proposed computational methodology that creates 2.5D or 3D representations of objects on the artwork so that it enabled PVI to understand the object by its shape and depth. 
 While these tactile or vibration feedback can help PVI to experience artwork non-visually in addition to audio feedback, limitations still exist as these require PVI to have access to special devices or custom artifacts.    
 Thus, we decided to propose a system that allows PVI to have access to various artwork anywhere through their personal touchscreen-based devices such as smartphones instead of requiring travelling or specific devices.  

\subsection{Art Criticism Method}
\textit{Feldman Model of Criticism} \cite{PracticalArtCriticism} is the most famous and common method to criticise artwork among art educators due to its simplicity and clear objectives which makes people can easily remember and use the method on their own terms \cite{prater2002art}. There are several other methods \cite{mittler1994art, prater2002art} to criticise artwork that are devised based on Feldman's model with few minor modification. As described in Table~\ref{tab:feldman}, this model consists of only four steps ---\textit{Description, Analysis, Judgement, and Interpretation}--- and each step has several clear questions that enables us to know whether we precisely completed the step or not. Moreover, it does not require any former education or knowledge to understand itself. Anyone can easily criticise and appreciate artwork using this method. 
Since all of our participants barely have prior knowledge about art and most of them have not seen or heard of paintings we used for the user study, we chose Feldman's model to evaluate their ability to appreciate and criticise paintings using our system.

\section{Preliminary Study}\label{preliminary}

Prior to the development of an accessible online art gallery for PVI for providing greater access to a larger number and variety of artwork with the help of collective intelligence similar to Wikipedia, we conducted a preliminary study to investigate the feasibility of collecting descriptions of paintings from a general crowd who do not have expertise in art. 
%proposal and implementation of a crowd-powered service as an online art gallery
\subsection{Method}\label{data method}
\begin{figure}[!b]
  \includegraphics[width=\linewidth]{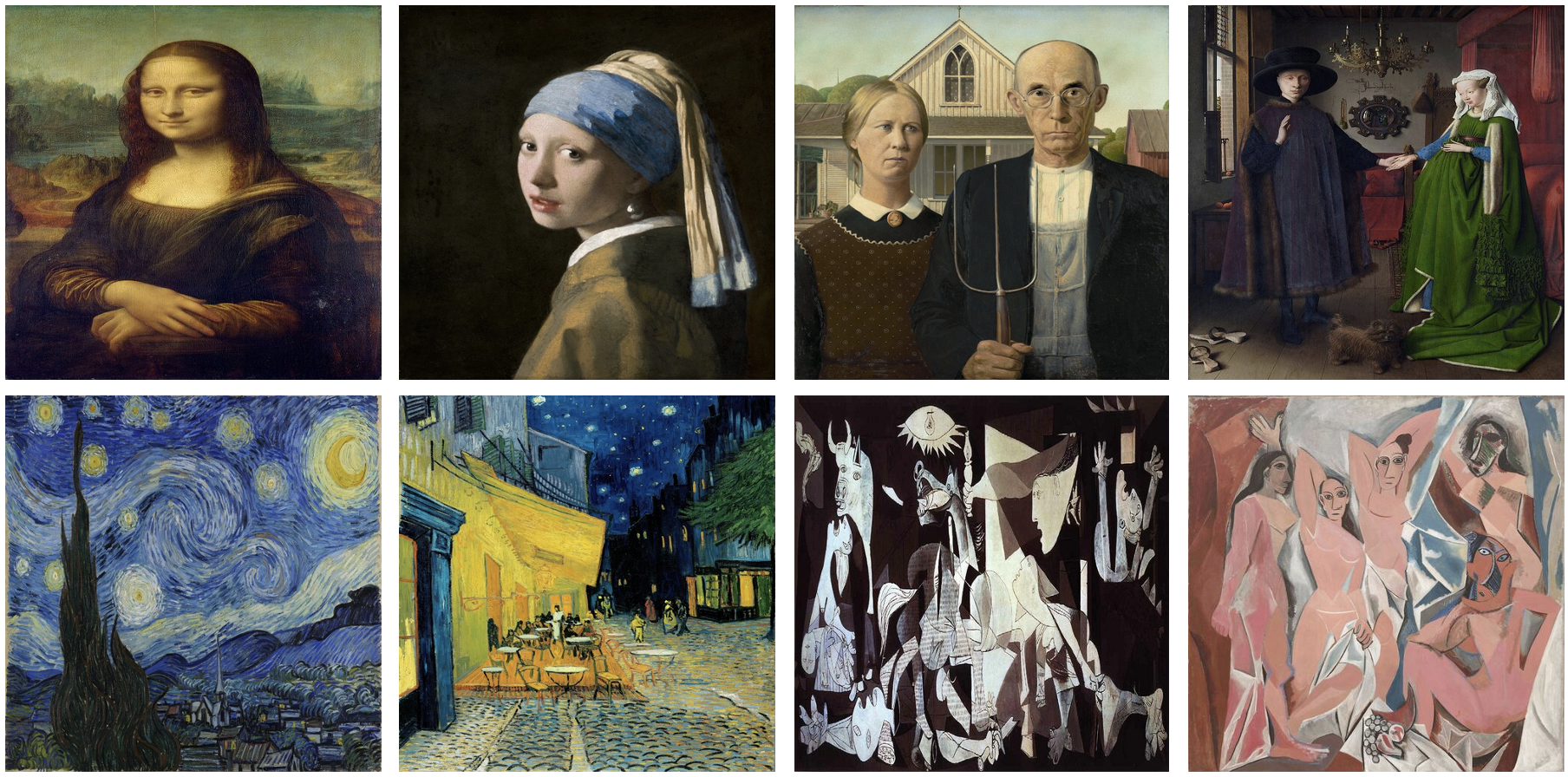}
  \caption{Paintings in our annotation tasks where the labels are numbered from top to bottom and from left to right:
  (A1) \textit{Mona Lisa} by \textit{Leonardo da Vinci}, 
  (A2) \textit{Girl with a Pearl Earring} by \textit{Johannes Vermeer}, 
  (A3) \textit{American Gothic} by \textit{Grant Wood}, 
  (A4) \textit{The Arnolfini Portrait} by \textit{Jan van Eyck}, 
  (A5) \textit{The Starry Night} by \textit{Vincent van Gogh},
  (A6) \textit{Cafe Terrace at Night} by \textit{Vincent van Gogh}, 
  (A7) \textit{Guernica} by \textit{Pablo Picasso}, and
  (A8) \textit{Les Demoiselles d'Avignon} by \textit{Pablo Picasso}. Note that the aspect ratio of each painting is adjusted to present all paintings in the same size. 
  }
  \label{fig:mturk_images}
\end{figure}

For this study, we designed an object-level annotation task of artwork and recruited crowd workers to perform the task. We used artworks for this task as shown in Fig.~\ref{fig:mturk_images}. We chose paintings intentionally in pairs varying compositions and genres to explore whether there are meaningful statistical similarities between paintings within the pair or not; A1 and A2 are portraits of a woman, A3 and A4 are portraits of a couple, A5 and A6 are landscape post-impressionism paintings by the same artist (\textit{i.e.}, Vincent van Gogh) and A7 and A8 are cubism paintings by \textit{Pablo Picasso}. 

\subsubsection{Apparatus}
\begin{figure}[t!]
\includegraphics[width=\linewidth]{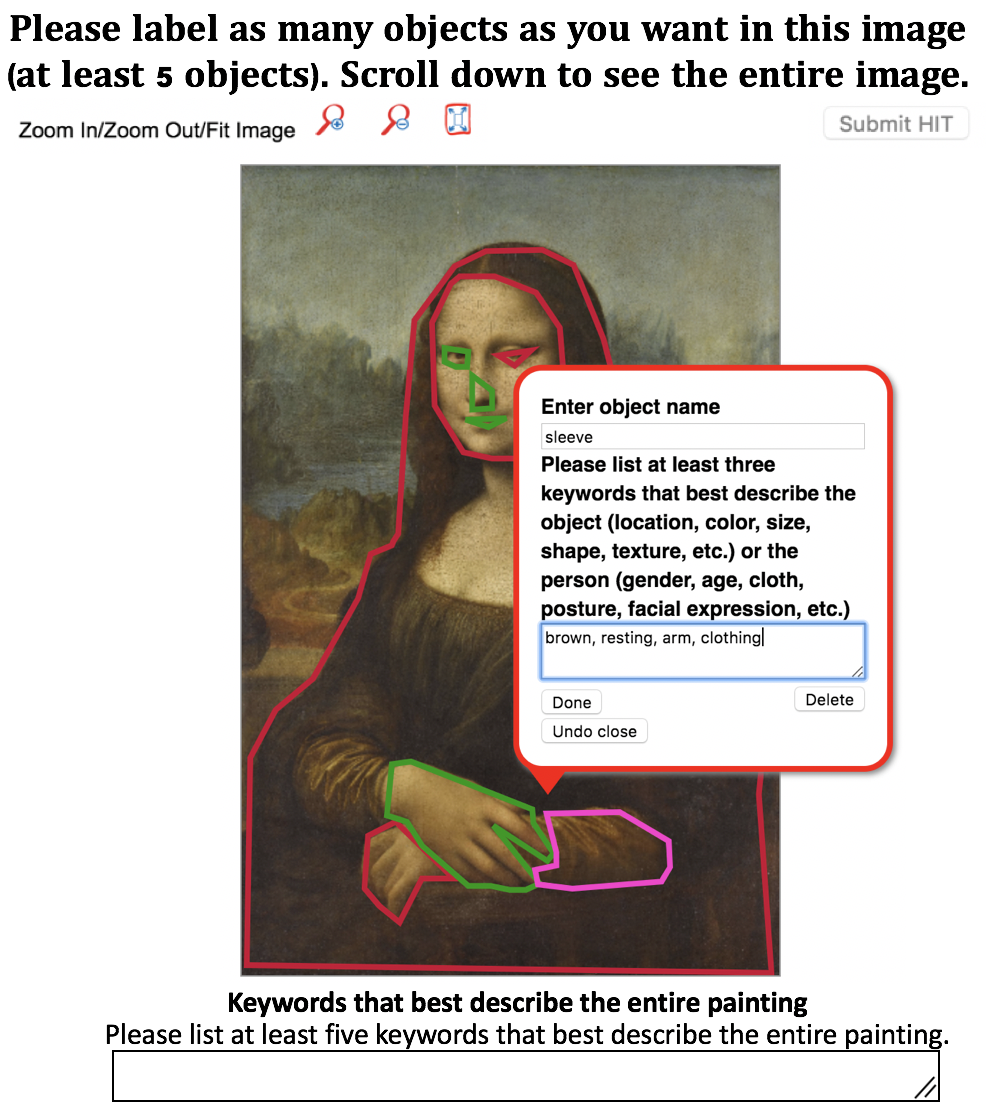}%
  \caption{Example screenshots of annotation pages used in our preliminary study.}
  \label{fig:mturk}
\end{figure}

To collect object-level information and overall description of each of the 8 paintings we chose for this study, we designed an instruction page as well as annotation page as shown in Fig.~\ref{fig:mturk}. As for the annotation page, we modified LabelMe \cite{russell2008labelme}, which is an online annotation tool to build image databases for computer vision research, so that crowd workers can provide not only the label but also additional information for each object that they have segmented and overall descriptions of the painting that they have worked on. In addition, we enabled \textit{Zoom} and \textit{Fit Page} features to help workers easily switch between magnified view and the original image. The interface was implemented through HTML, Javascript, and CSS on our Amazon EC2 server using Ubuntu to be posted for MTurk. It supports various web browsers including Chrome, Firefox, and Safari except for Internet Explorer.

\subsubsection{Data Collection Procedure}

To collect the object-level as well as painting-level descriptions from crowd workers, we posted our task to MTurk as a HIT (Human Intelligence Task) for each of the eight paintings. Workers who accepted our HIT were given an URL as a survey link of each HIT which would redirect them to our instruction page. The instruction page provides task explanations with visual examples in 3 steps: (1) segmenting objects, (2) labeling and providing descriptive keywords of the segmented objects and (3) providing descriptive keywords for the entire painting. 

\textbf{\textit{Step 1. Draw a shape around the object. }} 
First, workers were instructed to draw a polygon around an object directly on the presented painting on the screen to specify the boundary of the object with a series of clicks where each click defines a vertex of that polygon as below:
\indent\textit{\begin{itemize}
       \item Follow the boundary of the object it contains with clicks
       \begin{itemize}
           \item A straight line will show up connecting two successive clicked points
           \item Connect the last click point with the first one to complete a shape
       \end{itemize}
       \item If an object has several parts, please draw a shape for the parts as well as the whole object (e.g., face, eyes, nose, lips instead of just face/hat, umbrella instead of just person)
    \end{itemize}
}
Note that to collect as many object segments in each painting as possible while keeping the task as simple as possible as suggested by Zhang \textit{et al.}  \cite{zhang2011crowdsourcing}, we showed the segmentation results that were previously performed by others to every worker and asked them to perform the task for the rest of the objects that have not yet been segmented or objects that need to be improved and asked them to segment five objects rather than every object in each painting.

\textbf{\textit{Step 2. Enter object name and visual description of the object.}} 
When a worker finishes drawing a polygon of an object, a pop-up window shows up for the worker to enter the label of the object along with at least three keywords for providing visual descriptions as shown in Fig.~\ref{fig:mturk}b. Workers were also instructed to annotate whether the object is a part of the entire object or the object has several parts. The specific instruction we provided is written below. 
\indent
 \textit{
   \begin{itemize}
       \item Enter object name
       \item Enter a visual description of the object
       \begin{itemize}
           \item Please list at least three keywords that best describe the object (location, color, size, shape, texture, etc.) or the person (gender, age, cloth, posture, facial expression, etc.)
           \item Please note that there is no correct description. Please write down the words, however you see or feel
       \end{itemize}
       \item Press the `Done' button
       \item Enter a visual description of the object
    \end{itemize}
}
\textbf{\textit{Step 3. Providing descriptive keywords for the entire painting. }}
Finally, workers were asked to enter at least 5 keywords that best describe the entire painting as an option. Workers could start the task by entering their worker ID, which is an unique ID granted by MTurk, and click the \textit{`Submit'} button at the end of the instruction page, which redirects them to the annotation page.
Once workers are done with the annotation task, they can click the \textit{`Submit'} button at the very end of the annotation page to get a survey code which allows them to get the reward. 
All of our HIT for each painting was posted on July 27th, 2019 as a survey link project of MTurk. 
We allocated 20 workers per painting and limited the task execution time to be 30 minutes. 

Each worker was rewarded \$0.1 per HIT. Personally identifiable data was not collected at all throughout the procedure and the participation in this study was all voluntary.

\subsection{Findings}

%We collected crowd data through Amazon Mechanical Turk. 
To assess the feasibility of our approach for collecting object-level annotations of paintings from crowd workers, we analyzed our data mainly in terms of reliability and diversity. In addition, we examined the content of the keywords provided by workers to understand how we can better shape the instruction and the design of the annotation task to get informative descriptions in the future. 

\begin{figure}[t!]
  \includegraphics[width=1.0\linewidth]{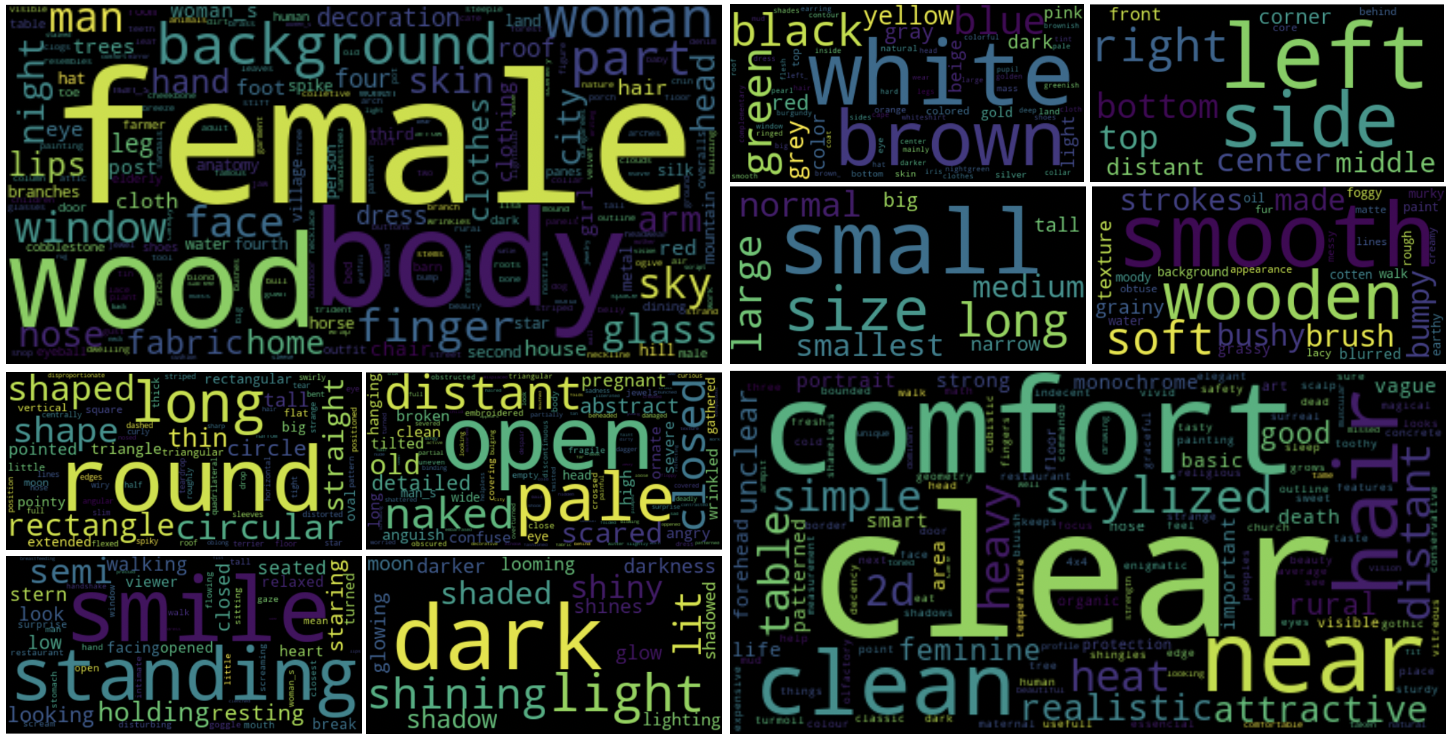}
  \caption{Pieces of word cloud visualization for each of the following ten categories of attributes collected across 8 paintings: object (top left), hue, location, size, texture, shape, status, pose/action, brightness (top left, top right, bottom left, bottom right at the top right and bottom left corners, respectively) and other (bottom right). The size of the font indicates the frequency of the keyword.}
  \label{fig:word cloud_attr}
\end{figure}

\begin{figure*}[b!]
\centering
  \includegraphics[width=0.8\linewidth]{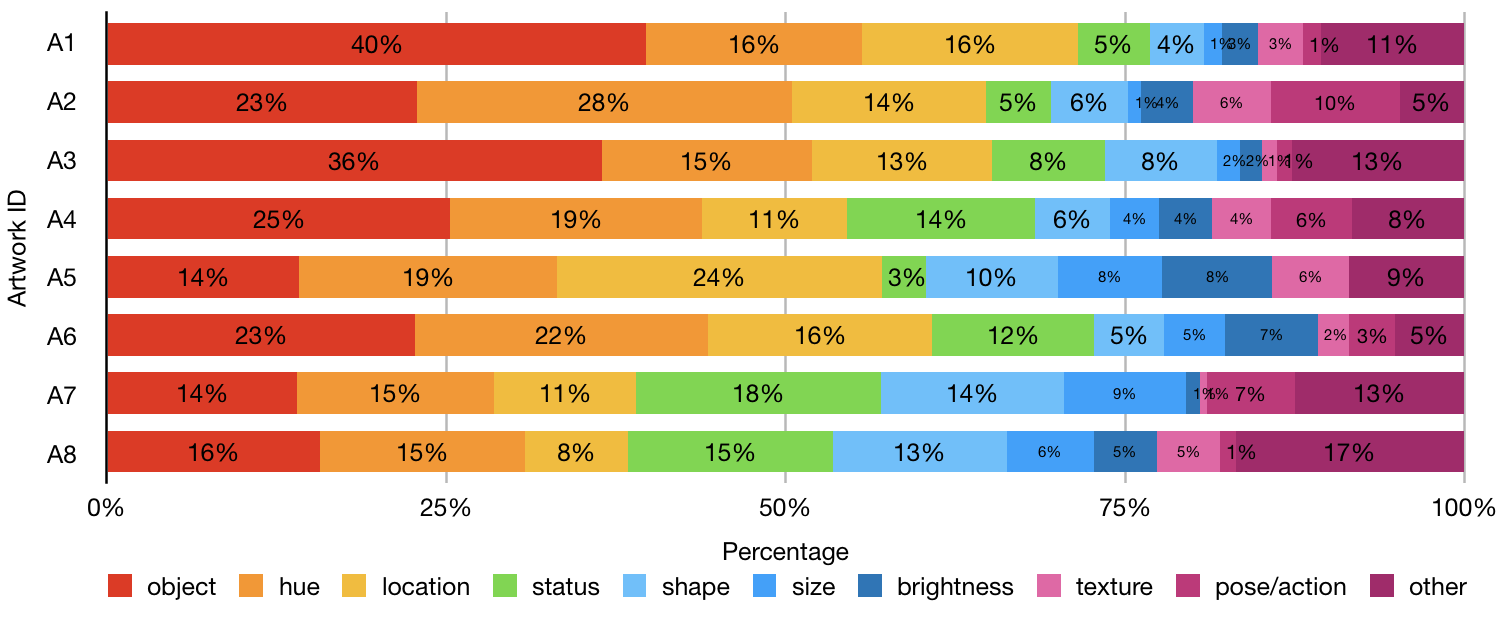}
  \caption{A stacked bar chart that shows the percentages for each of the attribution categories per painting.}
  \label{fig:percent_attr}
\end{figure*}

\subsubsection{Content Analysis of Object Annotations} 
%4개의 기준으로 나눈 속성 빈도수, 워드클라우드
\label{annotation analysis}

%\textit{Frequent Categories. } 
We intended to understand types of keywords provided by crowd workers for each object when no specific instructions were given; thus, we defined 10 categories. We first defined 3 categories which are \textit{hue}, \textit{location} and \textit{size}. Then we defined the rest as an \textit{expression} that does not belong to any of the first three categories and classified it into the following 7 subcategories: object (mostly another noun words that describe the segmented object), texture (\textit{e.g.}, soft, bushy), shape (\textit{e.g.}, thin, triangular), status (\textit{e.g.}, broken, scared), pose/action (\textit{e.g.}, staring, resting), brightness (\textit{e.g.}, dark, light), and other; see Fig.~\ref{fig:word cloud_attr} for more examples per category. 
%and checked whether it is in the lists of three categories. All keywords from attributes were sifted and stored in different lists of suggested 3 classes. 
For analysis, we developed a custom software written in Python that converts the annotation results in XML into a JSON format. Then, we trimmed every keyword to remove spaces. Afterwards, one of our internal researchers manually classified each keyword into one of the three categories if applicable. As for the rest, nouns and meaningless words like `is' or `not' were removed %from remains of the aforementioned filter function. For every remained keyword, 
and its parts of speech was checked using WordNet from Python NLTK package\footnote{http://www.nltk.org/howto/wordnet.html}. 
For all 8 paintings, the object keywords were the most frequent (23.2\%), followed by hue, location, status, other, shape, size, brightness, texture(18.0\%, 14.3\%, 10.4\%, 10.0\%, 8.3\%, 4.8\%, 4.2\%, 3.4\%, respectively.), and pose/action (3.4\%).

\begin{table}
\begin{tabular}{|l|l|l|l|l|l|l|l|l|}
\hline
                                    & \textbf{A1} & \textbf{A2} & \textbf{A3} & \textbf{A4} & \textbf{A5} & \textbf{A6} & \textbf{A7} & \textbf{A8} \\ \hline
\textbf{A1} &             & **          & \textit{n.s.}        & **          & ***         & **          & ***         & ***         \\ \hline
\textbf{A2} &             &             & ***         & \textit{n.s.}        & ***         & \textit{n.s.}       & ***         & ***         \\ \hline
\textbf{A3} &             &             &             & **          & ***         & **          & ***         & ***         \\ \hline
\textbf{A4} &             &             &             &             & ***         & \textit{n.s.}       & ***         & **          \\ \hline
\textbf{A5} &             &             &             &             &             & ***         & ***         & ***         \\ \hline
\textbf{A6} &             &             &             &             &             &             & ***         & ***         \\ \hline
\textbf{A7} &             &             &             &             &             &             &             & *           \\ \hline
\end{tabular}
\caption{Posthoc analysis of Chi-square tests for each pair of 8 paintings; `*' is for $p$ $<$ .05, `**' is for $p$ $<$ .01, `***' is for $p$ $<$ .001, and `\textit{n.s.}' is for results; that are \textit{not significant}.}
\label{table:annotation}
\end{table}

When examining the proportion of attribute categories per painting as shown in Fig.~\ref{fig:percent_attr}, portraits (A1-A4) seem to have higher percentages of object-related keywords although A6, which is not a portrait, is an exception. As for hue-related keywords, A2 has the highest proportion followed by A6, which indeed are the ones that are most colorful among 8 paintings. Interestingly, A5 has an exceptionally high percentage for location-related keywords. One possible explanation could be that the painting has almost identical objects (\textit{i.e.}, stars) repeated where their location is the most distinctive differences among them. While A5 and A6, which are paintings by the same artist, do not seem to have similarities in terms of the proportion, A7 and A8, two paintings by Pablo Picasso, seem to have some commonalities. For instance, these two paintings have relatively larger proportions of status- and shape- than object-related keywords compared to other paintings. Also, it has a large portion of \textit{other} category. This could be due to its cubism style as anything that are expressed are mostly abstract and vague. We conducted a Chi-square analysis of independence and found that the proportion of each category differs depending on the painting ({${\chi}^2_{(63)}$} = 263.82, $p$ $<$ .001, \textit{Cramer's V} = .16). As a posthoc analysis, we conducted multiple pairwise comparisons with Yates' correction using Python's SciPy library and the results are shown in Table \ref{table:annotation}. %revealed that A5 is significantly different from all other paintings ($p$ < .001). It has relatively large proportion of location-related descriptions which could be due to having the same object (\textit{i.e.} star) multiple times varying location. A7, a cubism-style painting, was also different from the rest of the paintings ($p$ < .001 for A1-6, and $p$ = .01 for A8). It has relatively high percentage of status-related keywords that described chaotic status of people and animals after a war in the painting. Similarly, A8, another painting by Pablo Picasso, was different from the rest ($p$ < .001 for A1-3\&6 and $p$ = .002 for A4) 

%A7 has a significantly different distribution of keyword categories compared to A1-5 ($p$ = .013 for A2 and $p$ < .001 for the rest). Similarly, the distribution of A8 was significantly different from A1 ($p$ = .002), A3 ($p$ = .001), A4 ($p$ < .001) and A5 ($p$ = .001). This suggests that A7 and A8, which are the two cubism-style paintings by Pablo Picasso, have relatively higher proportion of expression-, and size-related keywords and smaller proportion of location-related keywords. 

\begin{figure}[b!]
  \includegraphics[width=1.0\linewidth]{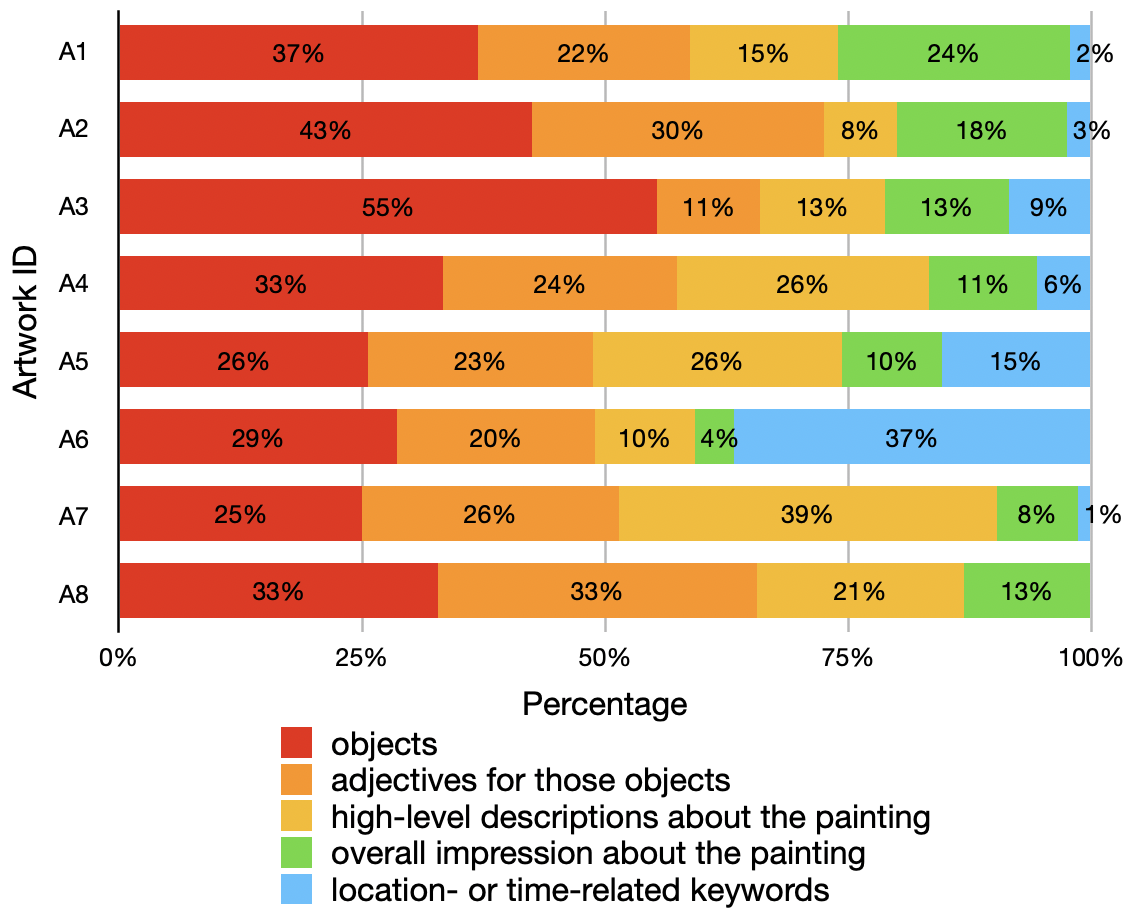}
  \caption{A stacked bar chart that shows the percentages for each of the five attribution categories of overall impressions per painting.}
  \label{fig:percent_impression}
\end{figure}

\subsubsection{Overall Impressions}\label{overall impression}
After removing repeated keywords within the keywords provided by the same worker, we collected 408 unique keywords from 106 valid sets across paintings. The average count was 51 (\textit{SD} = 11.1) per painting. We also examined overlapping keywords from different workers %shows the list of keywords that were provided by more than one worker per painting. 
and confirmed that only 153 out of 408 keywords were duplicates, which indicates that 75.7\% (\textit{N} = 309) of unique keywords across workers in total.   
We further examined the keywords (\textit{N} = 54) that appeared more than once and found that our workers had the tendency to list the labels or the attributes of the objects (\textit{e.g.}, smiling, old) that appeared in paintings when we asked them to enter keywords that best describe the entire painting. For example, keywords like a woman, girl, earring and couple were the keywords collected for portraits (A1-A4), while the moon, tree, stars were collected for landscape paintings (A5-A6). Moreover, these two paintings had a number of keywords that describe location or the time of the day (\textit{e.g.}, night) of the painted scene. Location-related keywords such as a cafe, outdoor, street were frequently observed especially for A6. %Also, keywords related to the overall atmosphere or personal impression of the paintings such as romantic, chaotic and weird seem to appear more common for paintings that are not portraits, which were A5-A6 and A7-A8. %Unlike the object-level keywords, it does seem easier to map those keywords to one of the paintings. 

We examined the percentages of each category of keywords; see Fig.~\ref{fig:percent_impression} for detail. Interestingly, in the case of extremely famous paintings like \textit{Mona Lisa} (A1), workers entered high-level descriptions about the painting such as the title or the name of the artist and that it is a famous art painting. Again, we conducted a Chi-square analysis of independence and found that the proportion of each category differs depending on the painting ({${\chi}^2_{(28)}$} = 110.13, $p$ $<$ .001, \textit{Cramer's V} = .26). A posthoc analysis show that paintings that are grouped together (\textit{e.g.}, A1\&A2, A3\&A4) are not statistically different. However, A6 is significantly different from A1, A3-A4 and A7-A8. The difference between either A7 or A8 and A2-A3 were also found to be significant.

\subsubsection{Descriptive Analysis}\label{descriptive analysis}

%Among 160 workers for all 8 images, we excluded 66 workers who did not complete any annotation at all (\textit{N} = 35) or submitted annotations with less than 5 objects (\textit{N} = 31) \footnote{Although we instruct workers to segment at least 5 objects, the task was set to submit the task regardless}. 
A total of 591 objects were annotated in total across 125 workers who has annotated at least one object for all 8 paintings while the rest 35 workers did not complete any annotation at all. However, we examined 528 of them which were annotated by 94 \textit{valid} workers who completed at least 5 annotation as instructed for the analysis\footnote{Although we instructed workers to annotate at least 5 objects, it was accidentally set to allow submitting the task regardless of the number of annotated objects.}. On average, 66 objects were collected per painting (\textit{SD} = 9.5) and the average number of annotated objects per workers was 5.6 (\textit{SD} = 1.3). 
In addition, the overall description per painting we collected at the end of the annotation task as an option, 117 descriptive keyword sets were collected across all paintings where the average number of the distinctive keywords per painting was 13.25 (\textit{SD} = 1.83). 
As for the duration, the time taken to receive 20 workers' submissions per painting from when HIT was uploaded was 10 hours on average (\textit{SD} = 2h 39m) where the average task completion time per worker was 11m 36s (\textit{SD} = 7m 13s).

\subsubsection{Data Reliability Assessment}\label{data reliability}
To assess the reliability of the annotation data collected by crowd, we examined each of annotations and checked for its validity. %All XML format data was converted to CSV format for more comprehensible checking process. 
Two researchers manually examined all the data and flagged invalid annotations given the following criteria: (1) an object is segmented as a single point rather than a polygon with at least three vertices or (2) a polygon is located at a wrong place for its label. 
As a result, the number of valid data was 408 out of 528, which accounts for 77.2\% of data collected by valid workers; five annotations by 3 workers did not meet both first and second criteria while 94 annotations by 24 workers, and 31 annotations by 16 workers were flagged in terms of the first and second criteria respectively. 

% 5개 이하 1개 이상으로 입력한 invalid worker 31명 의 경우 전체 63개 중에서 50개(76.37%)가 invalid. valid worker의 경우 528개 중 120개(22.73%) invalid
In addition, we examined validity of annotations created by 31 invalid workers who has annotated less than 5 objects. Among 63 annotations they completed, only 13 annotations which accounts for 20.6\% were valid. It is about 56.6\% lower than the validity rate of annotations from valid workers. We could find here that the data collected by whom had not followed the instruction is not reliable compared with the data from whom had followed it. 

The validity of the descriptive keyword sets describing each of the entire painting instead of objects was also checked; a researcher visually examined the sets and removed improper data (\textit{e.g.}, worker's unique ID) for MTurk or an URL of the task page. 
As a result, the number of valid sets for paintings was 106 out of 117 which accounts for 90.6\%. 

\subsubsection{Summary}
From this preliminary study, we confirmed the feasibility of collecting object-, and painting-level descriptions from crowd workers, although it requires validity check. In addition, we also identified that object, hue and locations are the most prevalent keywords that crowd workers provide when no specific instructions are given regardless of paintings. Moreover, we found that the workers tend to provide specific types of attributes more than others depending on the paintings for both object and painting level descriptions. Thus, specific guidelines should be provided to workers if one wish to have a balanced number of keywords across categories regardless of the painting type.  

\section{System Overview} 
%[시스템 설명 추가]
After the preliminary study, we designed and implemented AccessArt, which is a touchscreen-based online art gallery optimized for Apple's iOS devices with VoiceOver \footnote{https://www.apple.com/accessibility/mac/vision/} to enable PVI to appreciate a variety of artworks independently from anywhere using their own devices. As stated in the preliminary study, object-level descriptions of paintings were collected from crowd workers and used in a specific mode called \textit{Object Mode} of our system. In addition to the object-level exploration, we developed two extra modes---\textit{Overview Mode} and \textit{Part Mode}---to help PVI understand and appreciate the painting more easily. 

\subsection{The Design of the System}
\label{Interaction Mode}
Although our target users are PVI whose dominant feedback channel is non-visual, we designed three main modes (\textit{i.e., Overview Mode, Object Mode and Part Mode}) of interaction based on Shneiderman's Visual Information-Seeking Mantra~\cite{shneiderman1996eyes}. This suggests guidelines for designing information visualization applications, for conveying visual information of artwork to PVI while reducing cognitive overload; the three steps are \textit{overview first}, \textit{zoom and filter} and \textit{details on demand}. %In addition, we designed \textit{Edit Mode} so that anyone can contribute to the system by adding new information or editing existing content of paintings posted on our gallery similar to Wikipedia and WikiArt.

\begin{figure}[!b]
  \includegraphics[width=1.0\linewidth]{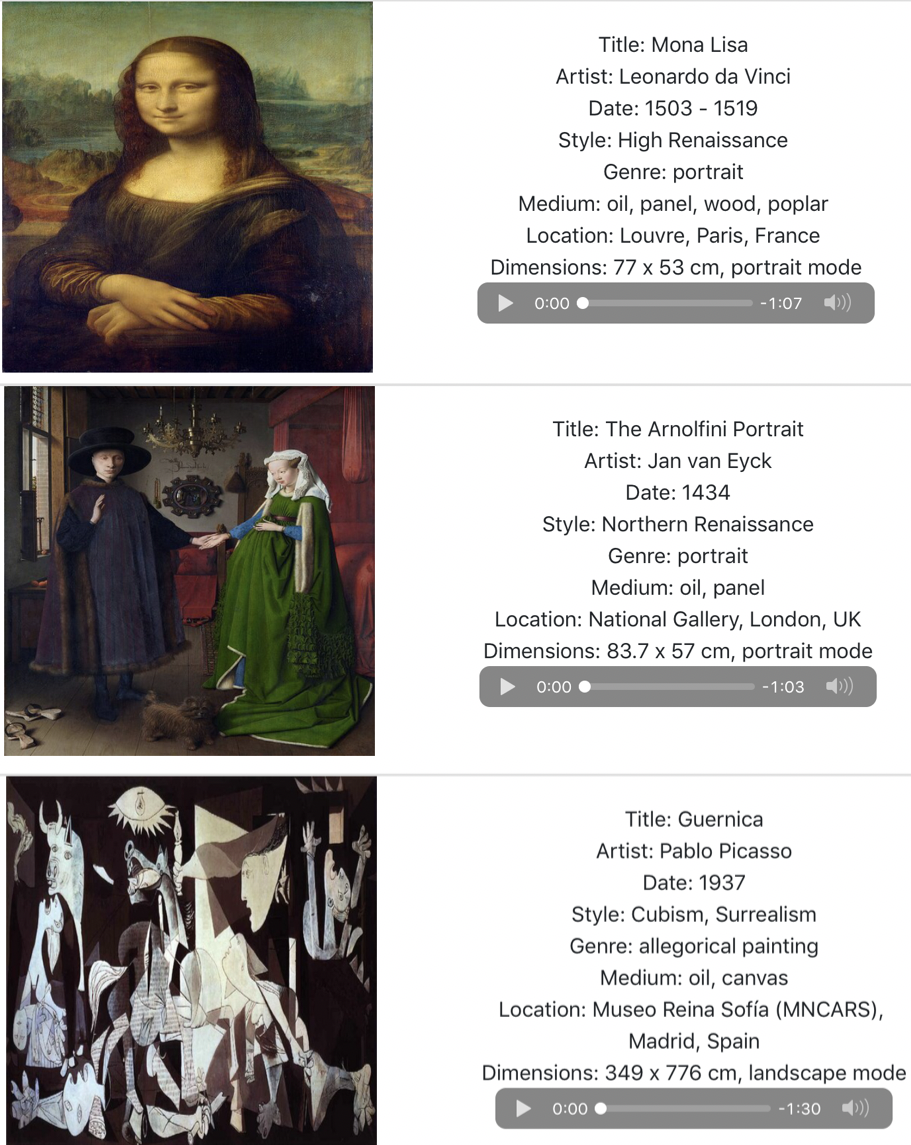}
  \caption{A screenshot example of \textit{Overview Mode} displaying \textit{Mona Lisa} by \textit{Leonardo da Vinci} (top; A1), \textit{The Arnolfini Portrait} (middle; A4) by \textit{Jan van Eyck}, and \textit{Guernica} by \textit{Pablo Picasso} (bottom; A7).%(left) and an appreciation page(right) of \textit{Mona Lisa} by \textit{Vincent Van Gogh} for an example. In the appreciation page, the image was optimized to fit to the mobile device's window size.
  }
  \label{fig:sys_overview}
\end{figure}

\subsubsection{Overview Mode}\label{overview}
We implemented this mode as the main screen of our system to provide users with an overview of paintings before they start exploring the paintings in depth. It is designed to help users to understand which paintings are presented in the gallery or listen to basic information of each painting. This basic information was extracted from WikiArt, such as a title, artist, year, and medium as shown in Fig.~\ref{fig:sys_overview}. The user can perform either swipe gestures anywhere on a touchscreen or drag their finger to navigate items on the screen while using VoiceOver. %If the user highlights the image, then a title, an ideal screen mode, and a overall description of the painting will be played through VoiceOver. 
We also informed users the ideal screen orientation (landscape \textit{vs.} portrait) of each painting. Moreover, users could listen to the overall description of a painting which contains the artists' painting style, the historical or social background when the painting was created, and widely accepted interpretation of the painting. 

\begin{figure}[t!]
\centering
\subfloat[\textit{Object Mode}]{{\includegraphics[width=7cm]{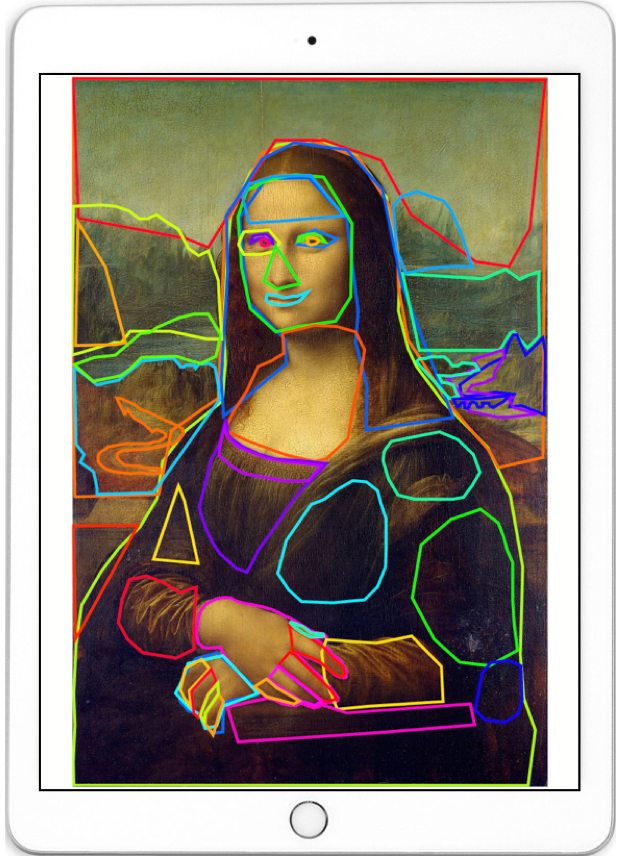}}}%
    \qquad
    \subfloat[\textit{Part Mode}]{{\includegraphics[width=7cm]{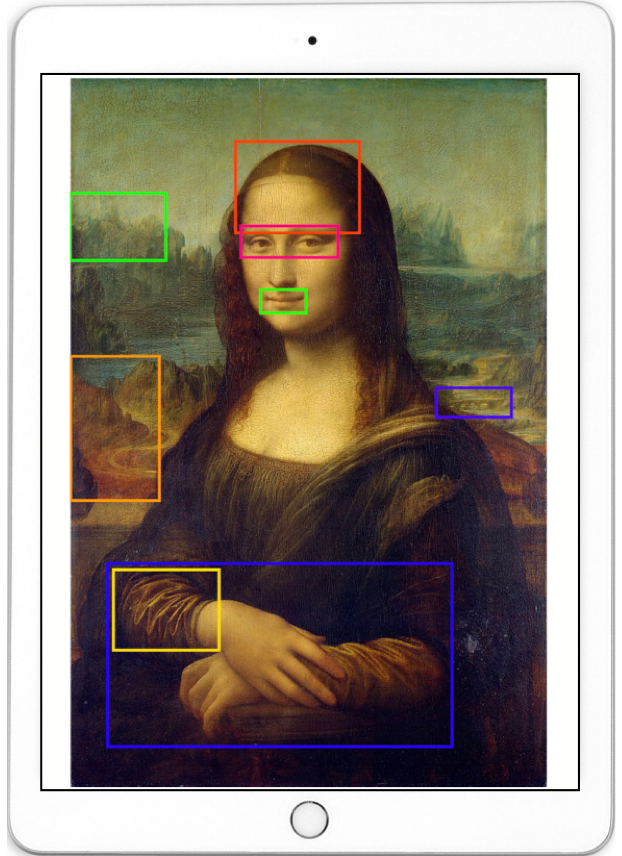} }}%
%    \caption{2 Figures side by side}
 % \includegraphics[width=0.9\linewidth]{figures/mturk_inst_task}
  \caption{Screenshot examples of (a) \textit{Object Mode} and (b) \textit{Part Mode}. Objects and part segments are shown in colored borderlines. In both modes, users can listen to verbal descriptions of each region upon touch.}
  \label{fig:object_mode}
\end{figure}

\subsubsection{Object Mode}\label{object_mode}
Users can navigate to this mode as default as shown in Fig.~\ref{fig:object_mode}a by selecting a certain painting from \textit{Overview Mode}. \textit{Object Mode} provides detailed object-level descriptions which were collected from crowd workers as stated earlier from the preliminary study in Section~\ref{preliminary}.
As supported from prior works \cite{morris2018rich, iranzo2019exploring}, this mode allows users to explore each object which is segmented in a polygon shape by touch while listening to object-level information in detail starting with its label followed by expression (\textit{e.g.}, object's texture, shape, and status. See Fig.~\ref{fig:word cloud_attr} for more examples), hue, location and size reflecting the frequency from the preliminary study. For example, if a user touches Mona Lisa's hair on the screen, the following verbal description will be played: \textit{``Hair. Dark, long, straight, smooth, shiny, attractive, female. The color is brown, black.''}. This allows users to focus on listening to the objects of interest instead of getting information of other undesirable objects at that moment. Moreover, the system provides an attribute filtering function to allow users to filter out unwanted information related specific types of attributes: expression, hue, location and size. For instance, the description example above is the result after filtering out location and size attributes.  

\subsubsection{Part Mode}\label{part mode}
This mode is designed to provide details upon users' request while or after exploring objects in \textit{Object Mode}. Users can perform split-tap (\textit{i.e.}, tapping with a second finger while holding one finger down) anywhere on the screen to switch modes between \textit{Object} and \textit{Part} modes. As shown in Fig.~\ref{fig:object_mode}, this mode highlights a set of regions and provides part-level descriptions in detail for each. These regions are suggested regions from a book called \textit{Great Paintings} that includes region-specific appreciation guides of 66 of the world's greatest paintings \cite{GreatPaintings}. For instance, the following is the description will be played when a user touches the lips of Mona Lisa in this mode:
%\begin{myquote}
\begin{quote}
 \textit{
 ``A mysterious smile: The smile of Mona Lisa is mysterious and soft, thanks to the Sfumato that has been applied to the painting and produced dramatic impacts. The woman's mouth is slightly raised, but it is almost impossible to read her facial expression. This smile also gives a tranquil atmosphere to the Mona Lisa."
 }
\end{quote}

\subsection{Implementation}\label{implementation}
Our application was originally developed as a web application so that it can be accessed from a web browser on any device with a URL link instead of downloading and installing the app. For implementation, HTML, JavaScript and CSS was used for designing the layout and styles, and  D3.js\footnote{https://d3js.org/} was used for supporting image-related visualizations and touchscreen interactions. 
Moreover, Amazon EC2 with Ubuntu was used to operate LabelMe to collect data from the crowd, and Firebase\footnote{https://firebase.google.com/} was also used to deploy our web application. The application was then optimized to a VoiceOver compatible mobile app for iOS devices. It supports various web browsers including Safari and Chrome.
%uran.oh: 중복 제거는 어떻게 했는지 자세히 적어주세요. 그리고 중복제거 후 최종으로 남은 데이터는 총 몇 개이고, 그림별로 또 몇 개인지.
Moreover, from the 408 valid data mentioned in Section~\ref{data reliability}, %we identified 120 invalid data and 408 valid data from collected annotations. For better results, 
we manually checked annotations once again for the system and removed 36 duplicated annotations that represent the same objects. 
Thus, for each painting, 46.5 objects were annotated on average (\textit{SD} = 7.8).

\section{User Study}
We conducted a user study with 9 participants with visual impairments to evaluate an effectiveness of our system for helping PVI better understand and appreciate paintings independently.

\subsection{Method}
We investigated the feasibility of our system with PVI by examining how well our system can support each of the four steps from \textit{Feldman Model of Criticism} \cite{PracticalArtCriticism}
%https://www.amazon.com/Practical-Criticism-Edmund-Burke-Feldman/dp/0137066740 
as presented in Table \ref{tab:feldman} 
, which is considered as an effective procedure for educating people to have better sense of criticizing and eventually appreciating artwork \cite{feldman1987varieties, anderson1988structure, anderson1991content}.

\begin{table*}
  \caption{Four steps of \textit{Feldman Model of Criticism}, and brief descriptions of each step.}
  \label{tab:feldman}
  \centering
  \begin{tabular}{llcc}
    \toprule
    \textbf{Steps}  & \textbf{Descriptions}\\
    \midrule
    Step 1. Description & Basic information and composition of the painting\\
    Step 2. Analysis & Integration of elements (style, structure, and technique, etc.)\\
    Step 3. Interpretation & Extracting meanings or messages from the painting\\
    Step 4. Judgment & Proposing a personal opinion based on prior steps\\           
  \bottomrule
\end{tabular}
\end{table*}

\subsubsection{Participants}
Nine PVI participated in our study. As summarized in Table \ref{participant}, 6 of them were male and their age is 36.2 on average (\textit{SD} = 7.7). Seven were totally blind and the other two had light perception; their onset years were varied. All participants were familiar with screen reader for mobile touchscreens (\textit{e.g.}, Apple's VoiceOver), and had prior experience of appreciating artwork in ways of visiting art exhibition. %Eight participants had plenty of interests in artwork. 
Participants were recruited through %a national school for the blind, and the others were recruited through %Siloam 
local communities for people with visual impairments. A reward worth about \$30 were given to each participant as compensation.

\begin{table*}[!t]
\setlength{\abovecaptionskip}{4pt}
  \small
  \caption{Participants' demographics, including age, gender and visual acuity of their best eye.}
  \label{participant}
  \centering
  \begin{tabular}{ccccc}
    \toprule
     PID & Age & Gender &  Visual Impairment (best eye) & Onset Years \\
    \midrule
    1& 48& Male & Totally blind& 10  \\
    2& 35& Male & Totally blind& 2 \\
    3& 32& Male & Totally blind& 30  \\
    %4& 54& F & low vision& 10& N\\
    4& 35& Male & Totally blind& 7 \\
    5& 30& Male & Light perception only& 8 \\
    6& 32& Female & Totally blind& 18\\
    7& 26& Female & Totally blind& 3\\
    8& 40& Female & Light perception only& 10 \\
    9& 48& Male & Totally blind& 10 \\
    
  \bottomrule
\end{tabular}
\end{table*}
%% 기존 10번 -> 4번

\subsubsection{Apparatus}
\begin{figure}[t!]
\centering
  \includegraphics[width=1.0\linewidth]{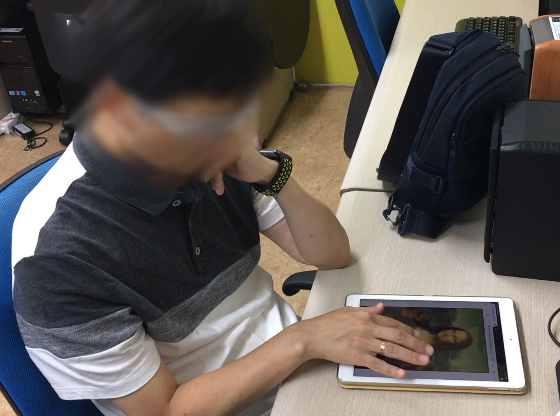}
  \caption{A participant (P9) exploring objects in Mona Lisa using our application in \textit{Object Mode}.}
  \label{fig:labsetting}
\end{figure}

We deployed a mobile web app version of our system described in Section \ref{Interaction Mode} prior to running the study  %As for \textit{Overview Mode}, among 8 paintings we developed, 
and accessed the link on an Apple's iPad 5th generation with a 10" screen during the study. We asked participants to sit on a chair in front of a desk and we placed the tablet on top of the desk as shown in Fig.~\ref{fig:labsetting}. Every session was audio-recorded for further analysis.

\subsubsection{Procedure}
After signing a consent form, each session began with background questionnaires followed by a semi-structured interview on their experience with art. Then participants were asked to use our system on a tablet to explore, appreciate and criticise 2 paintings using \textit{Overview Mode}, \textit{Object Mode} and \textit{Part Mode} after a tutorial. 
 
For the tutorial, we presented \textit{Overview Mode} as shown in Fig. \ref{fig:sys_overview} and asked participants to navigate and listen to the brief information (\textit{i.e.}, title, artist, date, style) about the first painting on the top of the page, which was \textit{Mona Lisa} by \textit{Leonardo da Vinci} (A1 from Fig.~\ref{fig:mturk_images}) using VoiceOver gestures; participants were allowed to use both linear navigation with swipes and exploration by touch, however and whenever they like throughout the study. After listening to the information of \textit{Mona Lisa} on their own, %we asked a few question about it. For the image appreciation task, 
we presented the painting on \textit{Object Mode} as shown in Fig. \ref{fig:object_mode}a to explore the painting and asked participants to answer what objects are depicted in the painting. Then we introduced \textit{Part Mode} (Fig. \ref{fig:object_mode}b) to participants and have them practice switching between the two modes. We allowed them to use this mode to explore the painting as well.%for a while, and asked few questions which could be answered via \textit{Part Mode}.
% 필터링
Before presenting the main two paintings, we explained participants about our attribute filtering function as described in Section \ref{object_mode}, and asked whether they wish to filter any attributes or not then set filters accordingly.  

We then asked participants to use our system to listen to basic information using \textit{Overview Mode} and explore and appreciate two paintings using either or both \textit{Object Mode} and \textit{Part Mode}. The paintings we used for the study were \textit{The Arnolfini Portrait} by \textit{Jan van Eyck} (A4) and \textit{Guernica} by \textit{Pablo Picasso} (A7).  
While participants were appreciating paintings, they were allowed to spend as much time as they need. 
After the appreciation, they answered several questions which were constructed based on 4 steps of \textit{Feldman Model of Criticism} as described in Table~\ref{tab:feldman}. Also, they were allowed to use our system freely to answer questions. In addition, we let participants listen to the overall description of the painting from \textit{Overview Mode}, and provide their perceived similarities and differences between the description and what they had appreciated. Finally, participants provided their subjective opinions and personal preference for two paintings. As for the order of the presentation, participants with even PID were presented with \textit{The Arnolfini Portrait} first, while others began with \textit{Guernica}.

\subsubsection{Data and Analysis}
We collected participants' demographic information, prior experience with artwork and subjective responses to questions during and after performing the task using AccessArt. For qualitative analysis, we removed all personally identifying information and assigned a unique identifier to each participants then coded the transcribed responses into a set of themes focusing on how our system can support each step in the Feldman model.

\subsection{Findings}
We summarize our findings focusing how our system, AccessArt, can be used to support each of the 4 steps in Feldman’s model of criticism to help people with visual impairments to explore and appreciate artwork independently and efficiently. %We also assessed 

\subsubsection{Supporting Description Step}
All 9 participants could easily describe the basic information (\textit{i.e.}, title, artist, and dimensions) and the composition of paintings using all three modes of interaction. 
They used \textit{Overview Mode} to get objective information about painting such as its title and artist.  

Then they tended to use \textit{Object Mode} to comprehend components of the painting first and changed the mode to \textit{Part Mode} if they wished to get more detailed information. 

While none of the participants knew \textit{The Arnolifini Portrait} (A4) and only 3 participants (P2, P4-5) had seen \textit{Guernica} (A7) before they lost their sight, most of the participants did not have any difficulty in listing objects depicted in the painting. 
For \textit{The Arnolfini Portrait}, all participants could easily list a man, a woman, and a dog which are main subjects of the painting when they were asked to describe which objects were drawn in the painting. Also, when the same question was asked for \textit{Guernica}, 2 participants (P1, P8), 
in addition to the ones who were aware of the painting, were able to list several objects such as broken body parts and animals based on information they obtained from \textit{Object Mode}. Moreover, other participants (P1, P3, P6-7, P9) were convinced that the painting is related to a war based on the attributes from object-level descriptions such as ``confused" and ``shocked", even though the word `war' was not directly mentioned at all. For example, P5 responded:

\begin{quote}
    \textit{``There are wailing mother and child in the painting. All of them have covered by flame in a war situation. I think a horse and a bull are emphasizing the sadness of the war."}
\end{quote}

\subsubsection{Supporting Analysis Step}
Participants were able to understand the arrangement of the painting (\textit{e.g.}, absolute or relative location of objects in the painting) and the overall color tone after exploring paintings with \textit{Object} and \textit{Part} modes. While most of the participants could answer these questions, the perceived difficulty seems to be different depending on the painting. 
% 구도
For instance, as for the arrangement of \textit{The Arnolfini Portrait}, all participants could easily describe the arrangement of the painting although P3 was a bit confused as %felt little confusion in imagining arrangement of two people from \textit{The Arnolfini Portrait}, simply because 
he thought that the woman would be lying on the bed. 
On the other hand, as for \textit{Guernica}, 3 out of 9 participants (P1-3) reported that it was hard to understand the locations of objects. %although they could sense what is happening in the scene. For example, P1 was still be able to assume that it is a battleground. 
For instance, P2 commented that this painting is more confusing compared to other paintings because it is a cubist style of painting. P3 also responded that the painting is strange and not making sense because he thought that a body part should be below a face, but it was not in reality. 

Still, the majority of the participants (P4-9) could explain the arrangement of painting quite accurately. For example, P6 described the arrangement of the scene precisely as below:

\begin{quote}
\textit{``Arrangement of the painting is chaotic and [objects are] scattered into pieces. A man lies horizontally at the bottom of the painting. A face is on the left side, and a woman embracing her child is at the above of the man overlapping with cattle. On the right side, there is a man and a woman staring the other side."} 
%어지럽고 조각조각나있고 흩어져있는 배치는 아래에 남자가 가로로 누워있고 왼쪽 얼굴이 있고 , 왼쪽에 남자보다 위에 소와 겹쳐서 아이를 안고있는 여자가 있고, 오른쪽은 처다보는 여자와 남자가 있음(P6)
\end{quote}
% 색깔
%Since no participants filtered 'color' attributes out by filtering function before the image appreciation task, 
We also asked participants to describe colors used in the paintings. For \textit{The Arnolfini Portrait}, all participants assumed that the painting would be colorful, bright and warm. For \textit{Guernica}, on the other hand, six participants (P1-4, P6-7, P9) assumed that the overall color of the painting would be monotonous and dark which was correlated to the color of actual painting while the remaining two participants thought that the painting would be full of intense colors as they imagined blood and fire from a war.

\subsubsection{Supporting Interpretation Step}
We could confirm that participants were able to guess what intention(s) the artists might have had when they were working on the paintings. 
We asked participants to share their own impression after appreciating each painting and their understanding of artist's intention. %Unlike other questions from the first two steps, all participants did not feel difficult to answer these questions. 
For \textit{The Arnolfini Portrait}, the responses were similar; participants felt wealth, warmth, and marital affection. 
For \textit{Guernica}, all participants received strong impression from the painting. %which the artist wished to convey through the painting. 
Six participants (P2-3, P5-8) described that the artist's intention of creating the painting was to convey complete misery and despair of wars, meanwhile other three felt hope. For example, %P7 said that she felt explosive emotions in each object representing individual story, and 
P6 mentioned that she felt hope as she noticed a light and a candle expressed in the painting.

\subsubsection{Supporting Judgment Step}
Finally, we asked participants what they liked or disliked about each painting if any.%, and how successful each artist was in expressing their message with their creation. 
 For \textit{The Arnolfini Portrait}, most of participants favored the relationship between married couple and relaxation from wealth of the rich middle-class people (\textit{N} = 6). P7 assumed that the unknown person in the mirror is probably a god, based on the fact that there were elaborate sculptures and a candle which represent religious beliefs.  
For \textit{Guernica}, responses of participants were mostly negative when they were asked for things they like or dislike about the painting. Four participants (P1-3, P8) said that they did not have any favorite part. P2, in particular, responded that he disliked almost every part of the painting because he does not prefer dark and gloomy atmosphere of the painting. On the other hand, four participants (P5-7, P9) preferred expressions of the artist used as a mean to convey a certain message. P9 also mentioned that he favors the message of anti-war movements that the artist tried to convey through his painting. %

% 메시지가 강할수록 주제를 더 쉽게 파악한다. -> DISCUSSION

\subsubsection{Supporting Overall Description} \label{similarity score}
% 전체설명과 감상결과가 얼마나 일치하는가. 
After completing each of the four steps per painting, we played the overall description of the painting in \textit{Overview Mode}. Then we asked participants to rate their perceived similarity between their own appreciation and overall description in a 7-point scale where 7 is best and the average score was high; 6.2 for \textit{The Arnolfini Portrait} (\textit{SD} = 0.7) and 6.0 for \textit{Guernica} (\textit{SD} = 0.9). %Although an appreciation of a certain painting could not be defined as a formulaic answer, we tried to identify whether participants understand overall composition and atmosphere of the painting or not in some extent. 
P2, whose score was 7 for \textit{Guernica}, and 6 for \textit{The Arnolfini Portrait}, noted that the overall description was slightly different from his own appreciation since an art appreciation is inevitably subjective while acknowledging that overall atmosphere and intention of an artist was similar. 
On the other hand, P8 rated 5 for \textit{Guernica} because she found out from the overall description that the color of the painting was generally monotonous while she imagined a colorful image. 
This indicates that personal bias could be removed by listening to an objective description.% which could be happened to sighted people too. 

While all participants agreed that they like being able to access basic background information of paintings, several concerns were raised. Five participants (P1-2, P5-7), in particular, believed that listening to overall description can disturb them when they are trying to have their own interpretation of paintings, especially it contains subjective criticisms or interpretations. For example, P7 said: 

\begin{quote}
\textit{``People who can see can make their own assessment of paintings by visually appreciating the artwork. But for me, I'll have to accept the descriptions as it is without having a second thought even if the information is not reliable."} 
\end{quote}

\subsubsection{Supporting Object and Part Modes} \label{professional desc}
We asked participants to share their impression of \textit{Object Mode} and \textit{Part Mode} and all participants showed positive reactions unlike the overall description. In general, they reported that changing two modes whenever they want and using both modes at the same time were significantly helpful in understanding paintings. 
P8 reported that having both modes is way more effective because it enables her to access desirable information quickly rather than providing a long description combining two kinds of information in a single mode. 
Moreover, 6 participants (P1, P3-5, P7, P9) emphasized that these two modes help them to elevate their understanding of paintings. P1 reported that he felt little confusion at first because he was not used to our system, but he said that he could understand the painting far more faster than simply listening to the overall description. 
P4 also mentioned:
\begin{quote}
\textit{``It feels like touching the painting directly in 3D space, and communicating with the painting. I even felt that I am maneuvering the painting dynamically."}    
\end{quote}
%In addition, P5 commented that he could have a chance to appreciate paintings in detail rather than merely accepting descriptions that typically happens at art exhibitions before.
Indeed, when we asked our participants if they prefer overall description of paintings or object-, and part-level descriptions using our system, all nine participants chose AccessArt as their preferred medium of appreciating artwork. 

\subsubsection{Supporting Attribute Filtering Function}

After the tutorial, we asked all participants whether they want to filter out any kind of object attributes and three participants turned off size-related information (P3, P7-8), and two participants turned off location-related information (P3, P6). Then, at the very end of the study, we asked participants to provide subjective feedback on this filtering function regardless of their usage of it. Seven participants (P1, P3-4, P5-8) provided positive responses. P1 reported that it is way more effective to use the filtering function because he can access desirable attributes more quickly without listening to redundant information which he can figure out by touching the paintings on the screen. P7 also said that it would be effective because desirable attributes are varied considerably for each person with a different degree of visual impairments. For example, she emphasized that size will be undesirable for people with moderate low vision. On the other hand, P9 said that filtering function is unnecessary as he would not filter out any information and instead try to get as much information as possible.

\subsubsection{Summary}
The findings with qualitative assessments demonstrated that PVI can understand paintings in terms of composition of the painting and the arrangements of components in the paintings from the object-level description produced by an anonymous crowd. Moreover, participants could capture the overall mood of paintings purely based on their own interpretation that they explored using our system yet similar to overall descriptions. 

\section{Discussion}

\subsection{The Feasibility of Collecting Data of Artwork from Crowd}

\subsubsection{The Reliability of the Crowd Workers and the Collected Data}

The HIT we designed for the preliminary study was set to automatically accept survey codes without checking for the validity of the codes themselves and the collected data. Thus, only about 60\% crowd workers have completed the task as instructed. Moreover, almost one fourth of the data was invalid even for the data produced by these workers.   
While we were still able to get a sufficient amount of reliable data, that we assumed to be more credible compared to annotations by a couple of researchers, to implement our system for supporting people with visual impairments for appreciating artwork with their own perspectives, a special consideration is needed to collect reliable data. To be scalable, in addition to relying on collective intelligence like Wikipedia, which is in fact our ultimate goal, various methods can be used such as a special incentive system to motivate workers to show high performance \cite{goodman2017crowdsourcing}. 
Similarly, Leimeister \textit{et al.} \cite{leimeister2009leveraging} also listed 4 motivations for crowd workers to provide more qualified data: learning, direct compensation, self-marketing, and social motives. We can try multiple methods to motivate workers such as informing them an objective or target users of the study. 
Besides, we confirmed that a rejection system that systemically excludes low quality results (\textit{e.g.}, accepting segmentation polygons that have at least 3 vertices) or a verification process that excludes invalid workers before they do the real task could be needed to collect qualified data as much as possible.  

% motivation으로 제대로 된 task를 하기를 기대할 수도 있고, 완전히 anonymous가 아닌 technical background가 있는 사람들에게만 받을 수도 있다. (MTURK 자체에서 지원) 혹은 task를 하기 전 연구의 목적이 무엇인지, 어떤 대상을 목적으로 하고 있는지 (시각장애인이 목적이다. 하는 정보를 명시)를 inform하면 더 나은 결과를 기대할 수도 있을 것이다. 
% 진짜 task를 시키기 전에 검증할 수 있는 간단한 task를 시켜 보는 것도 invalid worker를 사전에 차단하는 데에 도움이 될 것이다.
% multiple tasks for single worker 문제

%Although we have not assessed the effectiveness of \textit{Edit Mode} of our system that we implemented for this purpose to let anyone who wish to improve or correct any misleading contents or add missing information similar to web services based on collective intelligence such as Wikipedia. 

\subsubsection{The Efficiency of the Describing Objects and Paintings}
While our reward was relatively higher than 90\% of the MTurk HITs \cite{ipeirotis2010analyzing}, we could have divided into a series of simpler and constrained subtasks as suggested in a prior study \cite{zhang2011crowdsourcing} given that the average task completion time was over 10 minutes. 
For instance, we can design the task so that workers can draw a box around an object instead of drawing a polygon along with the object's boundary. Also, we can have three tasks, one for an object segmentation, one for object description and one for scene/atmosphere description. Moreover, as for the object description, while we intentionally allowed crowd workers to freely list any keywords to identify the types of attributes they crowd workers tend to provide, it would be more efficient to collect keywords for certain types of attributes. For example, we can prioritized types of attributes based on the needs of people with visual impairments who wish to learn about artwork; collecting expression-, and color-related information rather than keywords related to location and size. In addition, as color, location and size of an object can be computed automatically once the boundary of an object is specified, focusing on other general expressions that visually describes each object such as texture, shape and status while considering the genre of the paintings where one type of attribution is more informative than the others.    

Meanwhile, as for the description of the entire painting, the instruction needs to be more specific so that the data collected is more useful to people with visual impairments. It would be interesting to investigate if workers are more likely to provide informative annotations if they are informed that their data will be used for people with visual impairments for appreciating artwork.

\subsection{The Feasibility of Supporting the Appreciations of Artwork for PVI}

\subsubsection{Understanding Artwork from Object-level Descriptions}
There is no right answer for artwork appreciation and thus there is no objective measurement of how successful one has appreciated artwork. Thus, we asked questions based on 4 steps of \textit{Feldman Model of Criticism} focusing not on how closely each participant has understood general descriptions written by professional art critics, but on how participants were able to construct their own interpretation and judgment. 

In this regard, the participants from our study were able to understand the objects drawn in a painting, their arrangements as well as their attributes sufficiently well to capture the overall mood of a painting just from the object-level description. For instance, although the level of understanding can vary depending on the genre of the painting, 
even for \textit{Guernica}, which is a cubist style painting, participants could describe arrangements using our system. Moreover, they perceived that their impressions and interpretations were very similar compare to the overall description of a painting written by an expert. This suggests that our system can help people with visual impairments can access and assess paintings independently and creatively without passively relying on others' opinions. Still, we recommend providing the overall description in addition to object-level descriptions of artwork for the ones who wish to get more information including other people's interpretations. 
As a result, although verifying an ability of people to understand the artwork is difficult, we suggested the method to judge the ability of PVI to independently interpret or even criticize the paintings as an extension of prior works \cite{asakawa2018present, rector2017eyes}.

\subsubsection{The Effectiveness of Conveying Visual Information with Our Design}

As described in Section \ref{Interaction Mode}, we applied Shneiderman's Visual Information-Seeking Mantra when designing three main modes of interaction.  
Participants valued an ability to listen to the brief description of an object starting with its label followed by its attributes by scanning the painting on the screen as an overview in \textit{Object Mode} and switch the interaction mode to \textit{Part Mode} to get additional and detailed information on demand. In addition, the design also meets \textit{focus+context} \cite{card1999readings} as users can understand the components and the arrangements of objects in the depicted scene of the painting. 
Moreover, participants showed positive reaction towards attribute filtering function as it allows them to focus on information they wish to get and filter out the rest.
It would be interesting to investigate how the same approach can be applied for other types of artwork such as sculptures or in 3D space for describing the visual information about surrounding elements. 

\subsubsection{Applying a Vision-Based Method to Non-Visual Perception}
According to Arnheim \cite{arnheim1990perceptual}, despite absence of vision, sightless people can access and perceive artwork as sighted people with the help of haptic exploration, kinesthetical perception, and auditory information as an alternative way of vision. He claimed that 
In this regards, although Shneiderman's Visual Information-Seeking Mantra is designed to convey visual information for sighted people, we believe that this approach could also be effective for conveying the visual information of artwork in a non-visual way (\textit{i.e.}, verbally, kinesthetically, and haptically). 

Through the user study, participants could understand artworks as much as sighted people using our system. It insists that the vision-based method could also be applied for PVI in non-visual way to understand artwork.

\subsection{Limitations and Future Work}
Although the scope of study is to examine the feasibility of our approach for supporting people with visual impairments to appreciate artwork based on object-level description collected by the crowd, ours have limitations. First, as a feasibility study, a further investigation is needed for thorough evaluation of the effectiveness our system. For instance, we did not control crowd workers prior knowledge of the since eight paintings we have used for the study. Thus, the segmentation results may differ. While assumed that no workers have expertise in artwork, providing detailed instruction (\textit{i.e., ``Please list at least three keywords that best describe the object(location, color, size, shape, texture, etc.)''}), one's knowledge of specific artwork may have influenced the performance. 
Furthermore, we provided part-level descriptions which are from the book in \textit{Part Mode} which lacks scalability of our system.
As a future work, we should devise a more scalable method such as collecting professional part-level descriptions from art experts. 
We also plan to open our app to the public to examine if and how our system can be scaled up and maintained in the wild. Moreover, we wish to evaluate our system as an educational tool for anyone who are interested in the arts.

\section{Conclusion}
We investigated the feasibility of collecting visual descriptions of components in artwork from an an anonymous crowd based on annotation data of 8 paintings produced by Amazon Mechanical Turk workers. In addition, we proposed AccessArt, a web application system designed to enable people with visual impairments to appreciate various artwork online by listening to object-level descriptions drawn in a selected painting on their personal touchscreen-based devices then conducted a user study with 9 participants with visual impairments and collected their subjective feedback during and after using the proposed system to assess the potential in terms of how well our system could support each of the four steps in Feldman Model of Criticism. Our findings suggest annotations produced by non-expert crowd workers is capable of generating informative labels and attributes to help people with visual impairments to interpret and judge paintings independently based on their own perspectives rather than passively accepting the information written by professionals. As a future work, we plan to expand this work on a large-scale and make the system public to improve the accessibility of a greater number of paintings for people with visual impairments. \\

\begin{acknowledgements}
This work was supported by the Ewha Womans University Research Grant of 2018. Also, the research was supported by the Science Technology and Humanity Converging Research Program of National Research Foundation of Korea (2018M3C1B \\
6061353). 
\end{acknowledgements}
\small
\textbf{Compliance with ethical standards}
This study was approved in advance by a university Institutional Review Board (IRB) with the approval ID of ewha-201904-0001-01. 
\newline
\newline
\textbf{Conflict of interest} On behalf of all authors, the corresponding author
states that there is no conflict of interest.

% BibTeX users please use one of
\bibliographystyle{spbasic}      % basic style, author-year citations

\bibliography{sample-base}   % name your BibTeX data base

\begin{thebibliography}{43}
\providecommand{\natexlab}[1]{#1}
\providecommand{\url}[1]{{#1}}
\providecommand{\urlprefix}{URL }
\expandafter\ifx\csname urlstyle\endcsname\relax
  \providecommand{\doi}[1]{DOI~\discretionary{}{}{}#1}\else
  \providecommand{\doi}{DOI~\discretionary{}{}{}\begingroup
  \urlstyle{rm}\Url}\fi
\providecommand{\eprint}[2][]{\url{#2}}

\bibitem[{Anderson(1988)}]{anderson1988structure}
Anderson T (1988) A structure for pedagogical art criticism. Studies in Art
  Education 30(1):28--38

\bibitem[{Anderson(1991)}]{anderson1991content}
Anderson T (1991) The content of art criticism. Art Education 44(1):17--24

\bibitem[{Arnheim(1990)}]{arnheim1990perceptual}
Arnheim R (1990) Perceptual aspects of art for the blind. Journal of Aesthetic
  Education 24(3):57--65

\bibitem[{Asakawa et~al.(2018)Asakawa, Guerreiro, Ahmetovic, Kitani, and
  Asakawa}]{asakawa2018present}
Asakawa S, Guerreiro J, Ahmetovic D, Kitani KM, Asakawa C (2018) The present
  and future of museum accessibility for people with visual impairments. In:
  Proceedings of the 20th International ACM SIGACCESS Conference on Computers
  and Accessibility, ACM, pp 382--384

\bibitem[{Asakawa et~al.(2019)Asakawa, Guerreiro, Sato, Takagi, Ahmetovic,
  Gonzalez, Kitani, and Asakawa}]{guerreiro2019independent}
Asakawa S, Guerreiro J, Sato D, Takagi H, Ahmetovic D, Gonzalez D, Kitani KM,
  Asakawa C (2019) An independent and interactive museum experience for blind
  people. In: Proceedings of the 16th Web For All 2019
  Personalization-Personalizing the Web, ACM, pp 1--9

\bibitem[{Cantoni et~al.(2018)Cantoni, Lombardi, Setti, Gyoshev, Karastoyanov,
  and Stoimenov}]{cantoni2018art}
Cantoni V, Lombardi L, Setti A, Gyoshev S, Karastoyanov D, Stoimenov N (2018)
  Art masterpieces accessibility for blind and visually impaired people. In:
  International Conference on Computers Helping People with Special Needs,
  Springer, pp 267--274

\bibitem[{Card(1999)}]{card1999readings}
Card M (1999) Readings in information visualization: using vision to think.
  Morgan Kaufmann

\bibitem[{Cavazos~Quero et~al.(2018)Cavazos~Quero, Iranzo~Bartolom{\'e}, Lee,
  Han, Kim, and Cho}]{cavazos2018interactive}
Cavazos~Quero L, Iranzo~Bartolom{\'e} J, Lee S, Han E, Kim S, Cho J (2018) An
  interactive multimodal guide to improve art accessibility for blind people.
  In: Proceedings of the 20th International ACM SIGACCESS Conference on
  Computers and Accessibility, ACM, pp 346--348

\bibitem[{Clements(1979)}]{clements1979inductive}
Clements RD (1979) The inductive method of teaching visual art criticism.
  Journal of Aesthetic Education 13(3):67--78

\bibitem[{Feldman(1981)}]{feldman1987varieties}
Feldman EB (1981) Varieties of visual experience. Prentice Hall Englewood
  Cliffs, NJ

\bibitem[{Feldman(1994)}]{PracticalArtCriticism}
Feldman EB (1994) Practical art criticism

\bibitem[{Goncu and Marriott(2015)}]{goncu2015creating}
Goncu C, Marriott K (2015) Creating ebooks with accessible graphics content.
  In: Proceedings of the 2015 ACM symposium on document engineering, pp 89--92

\bibitem[{Goodman and Paolacci(2017)}]{goodman2017crowdsourcing}
Goodman JK, Paolacci G (2017) Crowdsourcing consumer research. Journal of
  Consumer Research 44(1):196--210

\bibitem[{Gyoshev et~al.(2018)Gyoshev, Karastoyanov, Stoimenov, Cantoni,
  Lombardi, and Setti}]{gyoshev2018exploiting}
Gyoshev S, Karastoyanov D, Stoimenov N, Cantoni V, Lombardi L, Setti A (2018)
  Exploiting a graphical braille display for art masterpieces. In:
  International Conference on Computers Helping People with Special Needs,
  Springer, pp 237--245

\bibitem[{Handa et~al.(2010)Handa, Dairoku, and
  Toriyama}]{handa2010investigation}
Handa K, Dairoku H, Toriyama Y (2010) Investigation of priority needs in terms
  of museum service accessibility for visually impaired visitors. British
  journal of visual impairment 28(3):221--234

\bibitem[{Hayhoe(2013)}]{hayhoe2013expanding}
Hayhoe S (2013) Expanding our vision of museum education and perception: An
  analysis of three case studies of independent blind arts learners. Harvard
  Educational Review 83(1):67--86

\bibitem[{Holloway et~al.(2019)Holloway, Marriott, Butler, and
  Borning}]{Holloway:2019:MSA:3290605.3300250}
Holloway L, Marriott K, Butler M, Borning A (2019) Making sense of art: Access
  for gallery visitors with vision impairments. In: Proceedings of the 2019 CHI
  Conference on Human Factors in Computing Systems, ACM, New York, NY, USA, CHI
  '19, pp 20:1--20:12, \doi{10.1145/3290605.3300250},
  \urlprefix\url{http://doi.acm.org/10.1145/3290605.3300250}

\bibitem[{Ipeirotis(2010)}]{ipeirotis2010analyzing}
Ipeirotis PG (2010) Analyzing the amazon mechanical turk marketplace. XRDS:
  Crossroads, The ACM Magazine for Students, Forthcoming

\bibitem[{Iranzo~Bartolome et~al.(2019)Iranzo~Bartolome, Cavazos~Quero, Kim,
  Um, and Cho}]{iranzo2019exploring}
Iranzo~Bartolome J, Cavazos~Quero L, Kim S, Um MY, Cho J (2019) Exploring art
  with a voice controlled multimodal guide for blind people. In: Proceedings of
  the Thirteenth International Conference on Tangible, Embedded, and Embodied
  Interaction, ACM, pp 383--390

\bibitem[{Karen Hosack~Janes(2012)}]{GreatPaintings}
Karen Hosack~Janes IZ Ian~Chilvers (2012) Great paintings

\bibitem[{Kwon et~al.(2019)Kwon, Koh, and Oh}]{10.1145/3308561.3354620}
Kwon N, Koh Y, Oh U (2019) Supporting object-level exploration of artworks by
  touch for people with visual impairments. In: The 21st International ACM
  SIGACCESS Conference on Computers and Accessibility, Association for
  Computing Machinery, New York, NY, USA, ASSETS ’19, p 600–602,
  \doi{10.1145/3308561.3354620},
  \urlprefix\url{https://doi.org/10.1145/3308561.3354620}

\bibitem[{Leimeister et~al.(2009)Leimeister, Huber, Bretschneider, and
  Krcmar}]{leimeister2009leveraging}
Leimeister JM, Huber M, Bretschneider U, Krcmar H (2009) Leveraging
  crowdsourcing: activation-supporting components for it-based ideas
  competition. Journal of management information systems 26(1):197--224

\bibitem[{Lichtenstein and Parker(2009)}]{lichtenstein2009wikipedia}
Lichtenstein S, Parker CM (2009) Wikipedia model for collective intelligence: a
  review of information quality. International Journal of Knowledge and
  Learning 5(3):254

\bibitem[{Low et~al.(2019)Low, McCamey, Gleason, Carrington, Bigham, and
  Pavel}]{10.1145/3308561.3354629}
Low C, McCamey E, Gleason C, Carrington P, Bigham JP, Pavel A (2019) Twitter
  a11y: A browser extension to describe images. In: The 21st International ACM
  SIGACCESS Conference on Computers and Accessibility, Association for
  Computing Machinery, New York, NY, USA, ASSETS ’19, p 551–553,
  \doi{10.1145/3308561.3354629},
  \urlprefix\url{https://doi.org/10.1145/3308561.3354629}

\bibitem[{{Metropolitan Museum of Art}(n. d.)}]{Metropolitan}
{Metropolitan Museum of Art} (n. d.) The audio guide by metropolitan museum of
  art. \url{https://www.metmuseum.org/visit/audio-guide}, accessed: 2019-10-14

\bibitem[{{Microsoft}(n. d.)}]{seeingAI}
{Microsoft} (n. d.) Seeing ai.
  \url{https://www.microsoft.com/en-us/ai/seeing-ai}, accessed: 2019-10-14

\bibitem[{Mittler(1994)}]{mittler1994art}
Mittler GA (1994) Art in focus. Glencoe

\bibitem[{Morris et~al.(2018)Morris, Johnson, Bennett, and
  Cutrell}]{morris2018rich}
Morris MR, Johnson J, Bennett CL, Cutrell E (2018) Rich representations of
  visual content for screen reader users. In: Proceedings of the 2018 CHI
  Conference on Human Factors in Computing Systems, ACM, p~59

\bibitem[{{Museum of Modern Art}(n. d.)}]{MOMA}
{Museum of Modern Art} (n. d.) Accessibility at museum of modern art.
  \url{http://www.moma.org/learn/disabilities/sight}, accessed: 2019-10-14

\bibitem[{{NeuroDigital Technology}(n. d.)}]{TouchingMP}
{NeuroDigital Technology} (n. d.) Touching masterpieces.
  \url{https://touchingmasterpieces.com/}, accessed: 2019-10-14

\bibitem[{Perkins(1994)}]{perkinsArts}
Perkins DN (1994) The intelligent eye: Learning to think by looking at art,
  vol~4. Getty Publications

\bibitem[{Prater(2002)}]{prater2002art}
Prater M (2002) Art criticism modifying the formalist approach

\bibitem[{Rector et~al.(2017)Rector, Salmon, Thornton, Joshi, and
  Morris}]{rector2017eyes}
Rector K, Salmon K, Thornton D, Joshi N, Morris MR (2017) Eyes-free art:
  exploring proxemic audio interfaces for blind and low vision art engagement.
  Proceedings of the ACM on Interactive, Mobile, Wearable and Ubiquitous
  Technologies 1(3):93

\bibitem[{Reinholt et~al.(2019)Reinholt, Guinness, and
  Kane}]{Reinholt:2019:ECE:3343055.3359722}
Reinholt K, Guinness D, Kane SK (2019) Eyedescribe: Combining eye gaze and
  speech to automatically create accessible touch screen artwork. In:
  Proceedings of the 2019 ACM International Conference on Interactive Surfaces
  and Spaces, ACM, New York, NY, USA, ISS '19, pp 101--112,
  \doi{10.1145/3343055.3359722},
  \urlprefix\url{http://doi.acm.org/10.1145/3343055.3359722}

\bibitem[{Rodrigues et~al.(2018)Rodrigues, Ferreira, Maia, Junior, de~Almeida,
  and de~Paiva}]{rodrigues2018image}
Rodrigues JB, Ferreira AVM, Maia IMO, Junior GB, de~Almeida JDS, de~Paiva AC
  (2018) Image processing of artworks for construction of 3d models accessible
  to the visually impaired. In: International Conference on Applied Human
  Factors and Ergonomics, Springer, pp 243--253

\bibitem[{Russell et~al.(2008)Russell, Torralba, Murphy, and
  Freeman}]{russell2008labelme}
Russell BC, Torralba A, Murphy KP, Freeman WT (2008) Labelme: a database and
  web-based tool for image annotation. International journal of computer vision
  77(1-3):157--173

\bibitem[{{Seattle Art Museum}(n. d.)}]{SAMA}
{Seattle Art Museum} (n. d.) Art beyond sight tours at seattle art museum.
  \url{http://www.seattleartmuseum.org/visit/accessibility}, accessed:
  2019-10-14

\bibitem[{Shneiderman(1996)}]{shneiderman1996eyes}
Shneiderman B (1996) The eyes have it: A task by data type taxonomy for
  information visualizations. In: Proceedings 1996 IEEE symposium on visual
  languages, IEEE, pp 336--343

\bibitem[{Stangl et~al.(2018)Stangl, Kothari, Jain, Yeh, Grauman, and
  Gurari}]{stangl2018browsewithme}
Stangl AJ, Kothari E, Jain SD, Yeh T, Grauman K, Gurari D (2018) Browsewithme:
  An online clothes shopping assistant for people with visual impairments. In:
  Proceedings of the 20th International ACM SIGACCESS Conference on Computers
  and Accessibility, ACM, pp 107--118

\bibitem[{{The Andy Warhol Museum}(n. d.)}]{OutLoud}
{The Andy Warhol Museum} (n. d.) Out loud, the andy warhol museum's inclusive
  audio guide.
  \url{https://itunes.apple.com/us/app/the-warhol-out-loud/id1103407119},
  accessed: 2019-10-14

\bibitem[{Winters et~al.(2019)Winters, Joshi, Cutrell, and
  Morris}]{winters2019strategies}
Winters RM, Joshi N, Cutrell E, Morris MR (2019) Strategies for auditory
  display of social media. Ergonomics in Design 27(1):11--15

\bibitem[{Zhang et~al.(2011)Zhang, Horvitz, Miller, and
  Parkes}]{zhang2011crowdsourcing}
Zhang H, Horvitz E, Miller RC, Parkes DC (2011) Crowdsourcing general
  computation. In: In Proceedings of the 2011 ACM Conference on Human Factors
  in Computing Systems, ACM, Association for Computing Machinery

\bibitem[{Zhong et~al.(2018)Zhong, Matsubara, and
  Morishima}]{zhong2018identification}
Zhong Y, Matsubara M, Morishima A (2018) Identification of important images for
  understanding web pages. In: 2018 IEEE International Conference on Big Data
  (Big Data), IEEE, pp 3568--3574

\end{thebibliography}
%\balanced

\end{document}